\documentclass[10pt,twoside,letterpaper,]{article}
\usepackage{amsmath}
\usepackage{textcomp}
\usepackage{mathtools}
\usepackage{siunitx}
\usepackage[margin=0.75in]{geometry}
\usepackage{algorithmicx}
\usepackage{authblk}
\usepackage{float}
\usepackage[font={footnotesize}]{caption}
\usepackage[hidelinks]{hyperref}
\author[1,2]{Jianhua Zhao}
\author[3]{Marcus A. Brubaker}
\author[1]{Samir Benlekbir}
\author[1,2,4,*]{John L. Rubinstein}
\affil[1]{Molecular Structure and Function Program, The Hospital for Sick Children}
\affil[2]{Department of Medical Biophysics, University of Toronto}
\affil[3]{Department of Computer Science, University of Toronto}
\affil[4]{Department of Biochemistry, University of Toronto}
\affil[*]{correspondence to: john.rubinstein@utoronto.ca}

\title{Description and comparison of algorithms for correcting anisotropic magnification in cryo-EM images}
\begin{document}
\maketitle
\begin{abstract}
\noindent Single particle electron cryomicroscopy (cryo-EM) allows for structures of proteins and protein complexes to be determined from images of non-crystalline specimens. Cryo-EM data analysis requires electron microscope images of randomly oriented ice-embedded protein particles to be rotated and translated to allow for coherent averaging when calculating three-dimensional (3D) structures. Rotation of 2D images is usually done with the assumption that the magnification of the electron microscope is the same in all directions. However, due to electron optical aberrations, this condition is not met with some electron microscopes when used with the settings necessary for cryo-EM with a direct detector device (DDD) camera. Correction of images by linear interpolation in real space has allowed high-resolution structures to be calculated from cryo-EM images for symmetric particles. Here we describe and compare a simple real space method, a simple Fourier space method, and a somewhat more sophisticated Fourier space method to correct images for a measured anisotropy in magnification. Further, anisotropic magnification causes contrast transfer function (CTF) parameters estimated from image power spectra to have an apparent systematic astigmatism. To address this problem we develop an approach to adjust CTF parameters measured from distorted images so that they can be used with corrected images. The effect of anisotropic magnification on CTF parameters provides a simple way of detecting magnification anisotropy in cryo-EM datasets.
 \end{abstract}

\section{Introduction}

\noindent Anisotropic magnification in electron microscope images of 2D crystals was first described more than 30 years ago \cite{baldwin1984measurement} but more recently has not been detected with most modern electron microscopes used under conditions typical for cryo-EM with film or charge-coupled device (CCD) cameras. Direct detector device (DDDs) cameras have revolutionized single particle electron cryomicroscopy (cryo-EM) \cite{kuhlbrandt2014resolution,smith2014beyond}. Two DDD manufacturers, Gatan and Direct Electron, have produced cameras with pixel sizes between 5.0 and \SI{6.5}{\micro\metre}, which is significantly smaller than is typical with CCD cameras. Because DDD cameras are placed below the projection chamber of the microscope, images formed on the DDD have an additional magnification relative to photographic film used with the same microscope. Consequently, electron microscopes used for high-resolution studies with a DDD may need to be set to a lower nominal magnification than was previously typical \cite{li2013electron}. At these conditions, anisotropic magnification has been detected in several microscopes, representing a large fraction of the instruments where this issue has been investigated. The phenomenon was seen first in a modern electron microscope with a 300 kV FEI Titan Krios microscope used with a Gatan K2 Summit DDD \cite{grant2015measuring} and has subsequently been detected on other FEI microscopes, including at least a 300 kV Tecnai Polara and a 200 kV Tecnai TF20. With a \SI{14}{\micro\metre} pixel size, the FEI Falcon series of DDDs does not require a low magnification setting and anisotropic magnification has not been detected with any microscope used in combination with this camera. \\

\noindent The consequence of anisotropic magnification in single particle cryo-EM is that images of molecules lying in different orientations on the specimen grid cannot be averaged coherently, limiting the resolution that can be obtained in 3D maps calculated from these data. For example, with 2 \% magnification anisotropy, a particle that is 300 \AA \ long would appear to be 300 \AA \ long when lying in one orientation and 306 \AA \ long when lying in another orientation. This effect will have more severe consequences for larger particles than for smaller particles. Anisotropic magnification can be detected by an elliptical appearance of the powder diffraction patterns calculated from images of a variety of specimens including polycrystalline gold, graphite, or thallous chloride. The observed anisotropy in magnification exists at the low magnification settings used to acquire images, but not at the higher magnifications used for correcting objective lens astigmatism. The cause of this effect has been proposed to be dirt in the microscope column that becomes charged and acts as an additional lens; in one 120 kV microscope the anisotropy was found to change slowly from 1.4 \% to 2.5 \% over the 7 years between 1983 and 1990 (Richard Henderson, personal communication). This observation, as well as our own measurements described below, suggest that anisotropic magnification is stable over the course of weeks or months, but should likely be measured periodically with any microscope. The dependence of magnification anisotropy on the nominal magnification of the microscope explains why the issue was not detected in current microscopes when higher nominal magnifications were typical. \\

\noindent In order to calculate high-resolution maps from cryo-EM images the contrast transfer function (CTF) of the microscope must be corrected. Most algorithms currently in use correct for the CTF during calculation of the 3D map. Introduction of anisotropic magnification after correction of objective lens astigmatism causes Thon rings from images to be elliptical, even when there is no objective lens astigmatism present. CTF parameters determined from elliptical Thon rings will therefore suggest the presence of astigmatism. A consistent ellipticity for Thon rings from images despite several attempts at correcting astigmatism at high magnification could indicate anisotropic microscope magnification. Similarly, the problem may be detected by the presence of systematic astigmatism in CTF parameters for datasets obtained over several EM sessions where different amounts of astigmatism are expected. An electron optical method for removing the effect has not yet been described. Here we describe and compare algorithms for computationally correcting the effects of anisotropic magnification on EM images. We also develop a method for recovering true CTF parameters from CTF parameters calculated from images with anisotropic magnification. The purpose of this manuscript is to increase awareness that the potential for anisotropic magnification exists, illustrate how this distortion can be detected and quantified, and demonstrate how it can be corrected computationally.\\ 

\section{Methods and Results}

\subsection{Measurement of magnification anisotropy}

\noindent To determine precisely the magnification of a FEI TF20 microscope operating at 200 kV and equipped with a Gatan K2 Summit direct detector, we acquired 99 movies containing crystalline thallous chloride particles. Anisotropic magnification parameters measured from these thallous chloride images can subsequently be used to correct cryo-EM images of biological specimens. The movies consisted of 30 frames collected at a rate of 2 frames sec\(^{-1}\), 5 e\(^{-}\)pixel\(^{-1}\)sec\(^{-1}\), and 1 e\(^{-}\) \AA\(^{-2}\) sec\(^{-1}\). From these movies, 558 thallous chloride particles that showed clear interatomic planes were selected (Fig. 1A). The thallous chloride lattice has a spacing of 3.842 \AA. Consequently, the power spectrum from each crystal is expected to show a spot at distance (\(p \cdot N\)/\SI{3.842}{\AA})  pixels from the origin, where \(p\) is the pixel size in \AA nstroms, and \(N\) is length in pixels along each edge of the image (Fig. 1B). The average of power spectra from images of particles is expected to produce a ring with this distance as the radius (Fig. 1C). However, the resulting average of power spectra had a slightly elliptical appearance. From \(4096 \times 4096\) averaged power spectra of images, each containing several thallous chloride particles, we carefully recorded the lengths and angles of 209 vectors, \(\mathbf{d}\), from the origin of the pattern to the diffraction ring. Fig. 1D shows a plot of the lengths of the vectors as a function of the angles they make with the positive \(k_x\)-axis of the power spectrum. The plot has a sinusoidal appearance, indicating that the diffraction ring from thallous chloride was indeed elliptical rather than round. Consequently, we fit the data to the equation

\begin{equation}
\label{ellipse}
|\mathbf{d}|=\frac{|\mathbf{r_{1}}|+|\mathbf{r_{2}}|+\cos (2 \cdot \left[{\theta_{d}-\theta_{r_1}}\right])\left(|\mathbf{r_{1}}|-|\mathbf{r_{2}}|\right)}{2}
\end{equation}

\noindent where \(|\mathbf{d}|\) is the distance from the origin of the diffraction pattern to a point on the elliptical diffraction pattern, \(\mathbf{r_{1}}\) and \(\mathbf{r_{2}}\) are the axes of the ellipse, \(\theta_{r_1}\) is the angle between the positive \(k_x\)-axis and \(\mathbf{r_{1}}\), and \(\theta_{d}\) is the angle between \(\mathbf{d}\) and the \(k_x\)-axis. The fit revealed that \(\mathbf{r_{1}}\) had a length that was 1.02 times the length of \(\mathbf{r_{2}}\) and is 1.3\textdegree \ from the \(k_x\)-axis of the pattern. \\

\begin{figure}[ht]
\centering
  \includegraphics[width=0.9\textwidth]{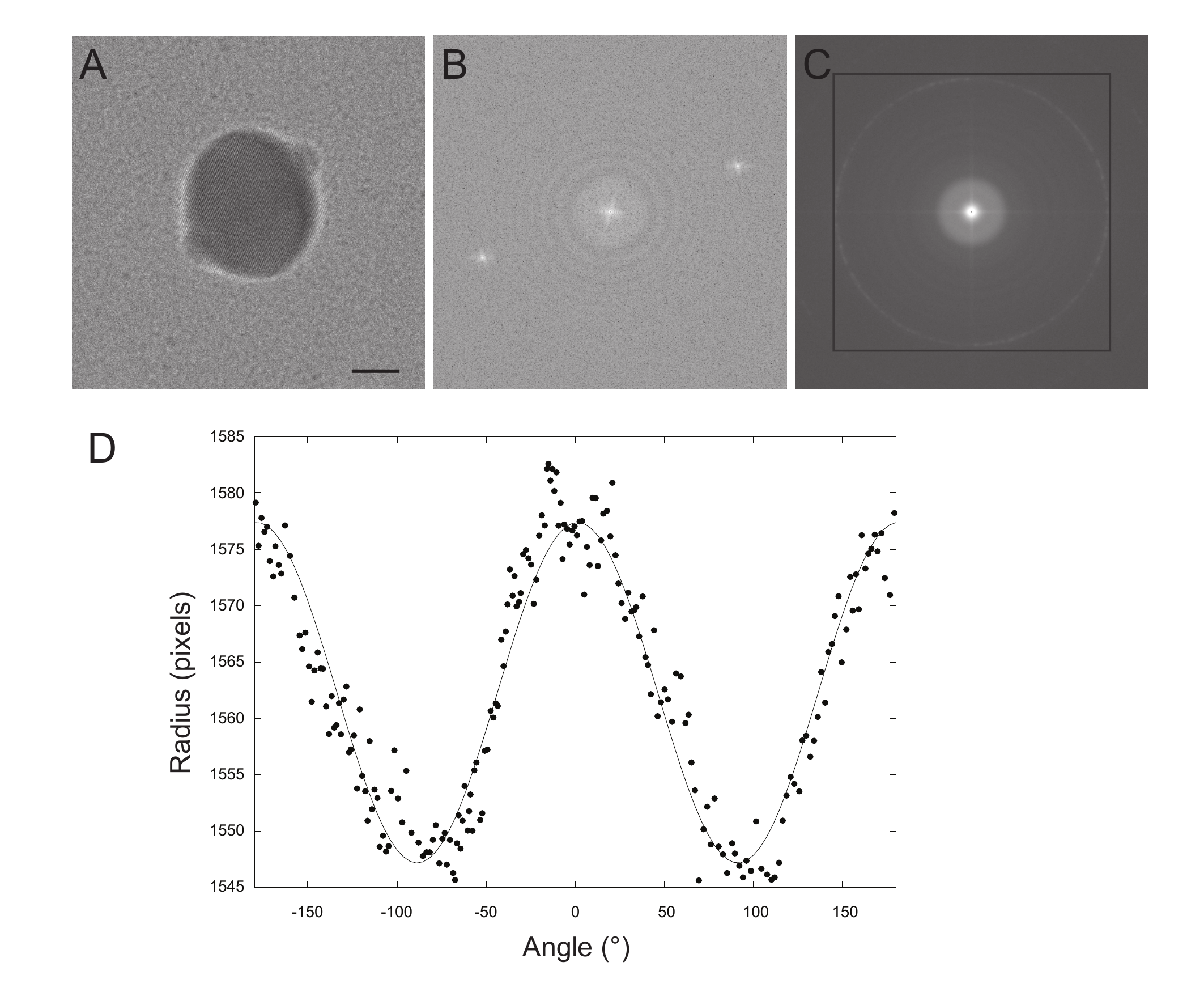}
  \caption{\textbf{Anisotropic magnification in images. A,} A thallous chloride particle in a 512 pixel \(\times\) 512 pixel image. Interatomic planes are apparent in the image. The scale bar corresponds to 100 \AA. \textbf{B,} The calculated power spectrum from the image in part A shows diffraction peaks. The expected distance from each peak to the centre of the image is \(\SI{1.45}{\AA} \cdot 512/\SI{3.842}{\AA}\) pixels. \textbf{C,} The average of 558 power spectra from thallous chloride particle images. The pattern is elliptical, not round, as seen by overlay of a perfect square that touches the pattern at its sides but not the top or bottom. \textbf{D, } A plot of the distance of the pattern from the origin of the power spectrum versus the angle of each point from the \(k_x\)-axis in the power spectrum from the average of 99 full frame power spectra padded to 4096 \(\times\) 4096 pixels.}
\end{figure}

\subsection{Correction of images for anistropic magnification}

\noindent Anistropic magnification can be corrected in images by appropriate stretching or contracting along \(\mathbf{r_{1}}\) or \(\mathbf{r_{2}}\). Stretching an image along one direction is equivalent to contracting its Fourier transform in the corresponding direction and vice versa. This equivalence is due to the similarity theorem \cite{bracewell1986fourier}, which states that if \(F(k_x)\) is the Fourier transform of \(f(x)\), then \(\frac{1}{a}F(\frac{k_x}{a}) \) is the Fourier transform of \(f(ax)\), where \(a\) is a constant and \(a \neq 0\). Consequently, four possible approaches were explored to correct images for anisotropic magnification: (1) A real space stretch along \(\mathbf{r_{2}}\) to make it equal to \(\mathbf{r_{1}}\), (2) a real space contraction along \(\mathbf{r_{1}}\) to make it equal to \(\mathbf{r_{2}}\), (3) a Fourier space contraction along \(\mathbf{r_{2}}\) to make it equal to \(\mathbf{r_{1}}\), and (4) a Fourier space stretch along \(\mathbf{r_{1}}\) to make it equal to \(\mathbf{r_{2}}\). The effect of stretching or contracting on the position of a point in an image or Fourier transform can be expressed in matrices as a transformation \(\mathbf{E}_{ani}=\mathbf{R}_{ani}\mathbf{S}_{ani}\mathbf{R}^\mathrm{T}_{ani}\), where  \(\mathbf{R^\mathrm{T}_{ani}}\) and \(\mathbf{R_{ani}}\) rotate points about the origin by the angles \(-\theta_{ani}\) and \(\theta_{ani}\), respectively, and \(\mathbf{S_{ani}}\) applies a stretch or contraction along the new \(x\)-axis. These matrices are defined by:

\begin{equation}
\label{anisotropy}
\mathbf{R}_{ani}=
\begin{bmatrix}
\cos{\theta_\mathit{ani}} & -\sin{\theta_\mathit{ani}}\\
\sin{\theta_\mathit{ani}} & \cos{\theta_\mathit{ani}}\\
\end{bmatrix}
\text{and }
\mathbf{S}_{ani}=
\begin{bmatrix}
a & 0\\
0 & 1\\
\end{bmatrix}.
\end{equation}

\noindent With these operations, the position of any point in the image \(\mathbf{p}=[x,y]\) or Fourier transform \(\mathbf{p}=[k_x,k_y]\) is transformed as \(\mathbf{p}^\prime=\mathbf{E_{ani}}\mathbf{p}\). Correcting anisotropic magnification requires applying the same transformation, but using a constant \(1/a\) where the microscope caused a distortion with magnitude \(a\). In each case, correcting anisotropic magnification requires interpolation because the positions of the transformed points of the image will in general not fall on previously sampled image points. Real space linear interpolation is a simple interpolation scheme (Fig. 2A). In linear interpolation in real space in one dimension, the value of the function \(f(x)\) is estimated from the nearest known values of the function, \(f(x_0)\) and \(f(x_1)\), as \(f(x)\approx \frac{x-x_0}{x_1-x_0}f(x_0)+\frac{x_1-x}{x_1-x_0}f(x_1)\). In other words, the unknown value of the function at the point \(x\) is treated as a weighted average of the known values of the function at the points \(x_0\) and \(x_1\). This approach can be extended two dimensions as bilinear interpolation, where \(f(x,y)\) is estimated from \(f(x_0,y_0)\), \(f(x_1,y_0)\), \(f(x_0,y_1)\), \(f(x_1,y_1)\) as
 
\begin{equation}
\begin{aligned}
f(x,y) \approx &\frac{x-x_0}{x_1-x_0} \cdot \frac{y-y_0}{y_1-y_0}f(x_0,y_0)+\frac{x_1-x}{x_1-x_0} \cdot \frac{y-y_0}{y_1-y_0}f(x_1,y_0)+\\
& \frac{x-x_0}{x_1-x_0} \cdot \frac{y_1-y}{y_1-y_0}f(x_0,y_1)+\frac{x_1-x}{x_1-x_0} \cdot \frac{y_1-y}{y_1-y_0}f(x_1,y_1).
\end{aligned}
\end{equation}

\noindent Numerous more sophisticated real space interpolation schemes may also be used. Bilinear interpolation can also be performed in Fourier space. Interpolation in Fourier space benefits from padding images with zeros before the Fourier transform to increase the sampling of the Fourier transform \cite{bracewell1986fourier}. In all of our Fourier space interpolation calculations we padded images to twice their original dimensions before performing the Fourier transform. Improved interpolation in Fourier space can be achieved with a \textit{sinc} interpolation approach (Fig. 2B). In Fourier space, each Fourier component on the regularly sampled lattice contributes to all off-lattice points as a normalized \textit{sinc} function or \(sin(\pi \Delta k)/(\pi \Delta k)\), where \(\Delta k = k-k_i\) is the distance between the off-lattice point and the on-lattice point in the Fourier transform \cite{bracewell1986fourier}. The value of the off-lattice point in a Fourier transform, \(F(k)\) is 

\begin{equation}
F(k)=\sum_{i=1}^{I}F(k_i)\mathit{sinc}(\pi \Delta k)
\end{equation}

\noindent where the summation is performed over all \(I\) on-lattice points. Summation over all Fourier components in an image for each interpolated point can be computationally expensive. Instead, an approximation of \(F(k)\) can be obtained by performing the summation over a limited set of points that are within some specified distance to the interpolated point. We elected to use on-lattice points that were 3 or fewer pixels away from the interpolated point as a compromise between accuracy and speed.\\

\begin{figure}[ht]
  \centering
  \includegraphics[width=0.9\textwidth]{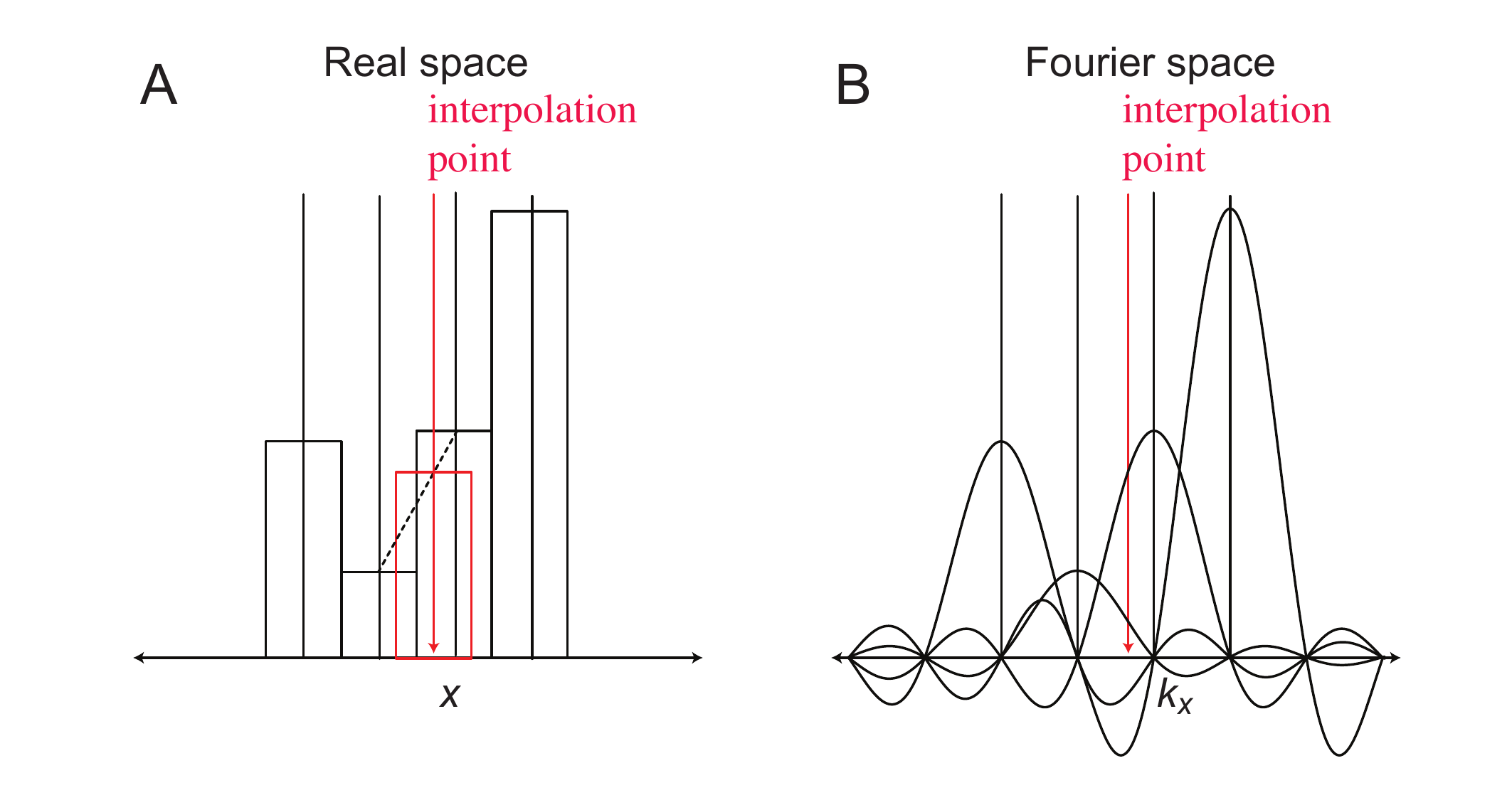}
  \caption{\textbf{Comparison of linear interpolation in real space and \textit{sinc} interpolation in Fourier space. A,} In real space the value of an off-lattice point is estimated by linear interpolation, where a weighted average of the nearest on-lattice points is obtained. \textbf{B,} In Fourier space, each on-lattice point produces a normalized \textit{sinc} function that is 0 at all on-lattice points and non-zero at all off-lattice points. A point is interpolated by adding all of the contributions from all of the on-lattice points or approximated by adding the contributions from nearby points. The performance of this \textit{sinc} function interpolation is improved by padding images with zeros before Fourier transforming in order to increase the sampling of the Fourier transform.}
\end{figure}

\noindent Every interpolation approach causes different artefacts in the image, the Fourier transform of the image, or both. Contraction in real space (approach 2) and stretching in Fourier space (approach 4) both would cause values from outside of the image to be pulled into the image frame. Creating artefacts at the edge of a real space image can complicate the process of floating images (setting the perimeter pixels to a mean of 0) and normalization based on the standard deviation of the image background. Consequently, we elected not to pursue approaches 2 and 4. Bilinear interpolation in real space (approaches 1 and 2) causes a banding pattern in the real space images, as shown for an image stretched in real space (Fig. 3Ai, red arrows). This artefact arises because the interpolated value is always less extreme than the values from which it is interpolated (Fig. 2A). Stretching of the image in real space also leads to some information being shifted to higher and lower frequency in the Fourier transform of the image (Fig. 3Bi, green arrows). This artefact is a result of the shifting of peak positions that occurs when performing linear interpolation. Contraction in Fourier space (approach 4) does not lead to any notable artefacts for the resulting image for either bilinear interpolation (Fig. 3Aii) or \textit{sinc} function interpolation (Fig. 3Aiii). It does, however, cause undefined values from outside of the Fourier transform to be pulled into the high-frequency edges of the Fourier transform (Fig. 3Bii and iii, blue arrows) but does not lead to any other noticeable banding artefacts in power spectra.\\

\begin{figure}[ht]
  \centering
  \includegraphics[width=0.8\textwidth]{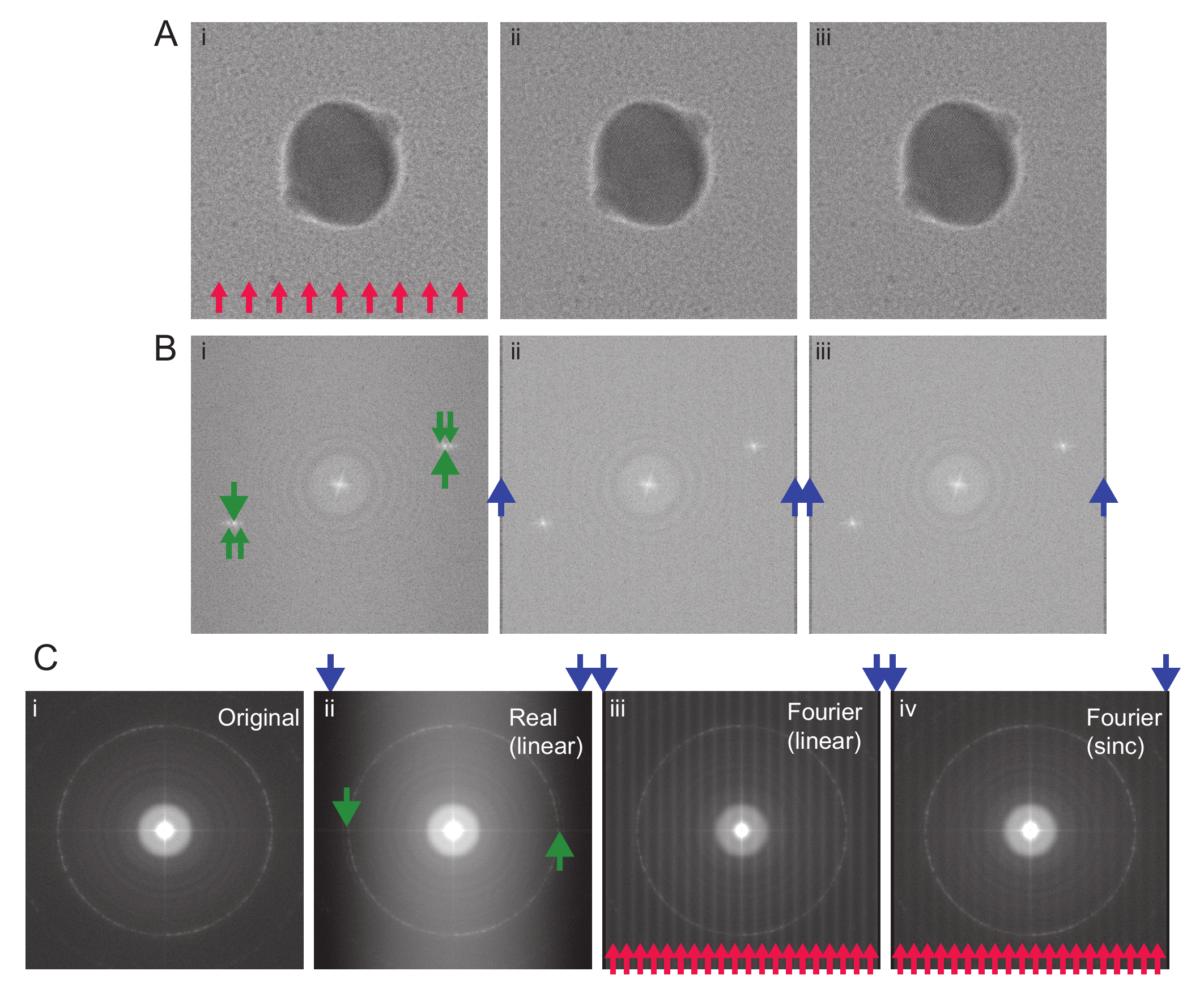}
  \caption{\textbf{Application of methods for correcting anisotropic magnification of images. Ai} Stretching images by 2 \% in real space causes a banding pattern artefact to appear on the image at a frequency of 2 bands per 100 pixels (red arrows). Stretching the image by contracting the Fourier transform with bilinear interpolation (\textbf{ii}) or \textit{sinc} interpolation (\textbf{iii}) causes no noticeable artefacts in the image. \textbf{Bi,} The real space stretch also causes part of the power of the diffraction peaks to be moved to higher and lower frequencies in the power spectrum of the image (green arrows). Contraction of the Fourier transform by either bilinear interpolation (\textbf{ii}) or \textit{sinc} interpolation (\textbf{iii}) requires that pixels be created to fill in the edge of the Fourier transform (blue arrows) but does not appear to cause any other notable artefacts in the power spectra from individual images. \textbf{Ci,} The average of power spectra from 558 uncorrected thallous chloride particle images shows no artefacts, other than the anisotropic magnification. \textbf{ii,} The average of power spectra from 558 thallous chloride particle images corrected for the 2 \% magnification anisotropy by stretching the real space image shows several severe artefacts, including a non-flat background in the power spectrum (blue arrows) and movement of information to higher and lower frequency in the direction of the correction (green arrows). \textbf{iii,} The average of power spectra from 558 thallous chloride particle images corrected for the 2 \% magnification anisotropy by bilinear interpolation to contract the Fourier transform reveals a strong banding pattern (4 bands per 100 pixels due to padding) that was not apparent from a single power spectrum (red arrows), as well as the edge pixels seen before (blue arrows). \textbf{iv,} The banding pattern is less intense when \textit{sinc} interpolation is used to contract the Fourier transform. }
\end{figure}

\noindent Inspection of the average of the 558 power spectra from thallous chloride particles is a sensitive way to detect systematic artefacts in power spectra. The average of power spectra from the original images does not have any apparent artefacts other than the anisotropic magnification (Fig. 3Ci). The average of power spectra from images corrected by stretching and bilinear interpolation in real space shows severe artefacts, such as a non-uniform background (Fig. 3Cii, blue arrows) and shifting of power from the diffraction peaks to both higher and lower frequencies in the direction of the correction (Fig. 3Cii, green arrows). The average of power spectra from images corrected by contraction in Fourier space shows a banding pattern from the finite inclusion of Fourier terms that is more intense for bilinear interpolation than for \textit{sinc} interpolation (Fig. 3Ciii and iv, red arrows). As a result of padding images before Fourier transform, the frequency of these bands is twice the frequency of the artefact bands in the real space interpolated image. The average of power spectra from images corrected by contraction in Fourier space also shows the edge effects from contraction seen in a single power spectrum (Fig. 3Ciii and iv, blue arrows). Consequently, we concluded that the small additional computational expense of the more sophisticated Fourier space \textit{sinc} interpolation scheme is justified. Undoubtedly, it would also be possible to improve the real space interpolation approach to the point where it matched the performance of the Fourier space \textit{sinc} interpolation approach: for example by padding the Fourier transforms of the images to increase the sampling density or by using cubic interpolation rather than linear interpolation. After correction of images by contraction of Fourier transforms, the powder diffraction ring from thallous chloride is found at a distance of 1546 pixels from the origin in the 4096 \(\times\) 4096 Fourier transform. This distance corresponds to a pixel size of 1.45 \AA \ and a magnification of 34,383\(\times\) on the DDD sensor.

\subsection{Effect on contrast transfer function parameters}

For high-resolution maps to be calculated from cryo-EM data, images must be corrected for the effects of the contrast transfer function (CTF). CTF parameters are typically measured from the power spectra of images and used to correct for the CTF during calculation of the 3D map. A stretch or contraction of an image causes an analogous change in the image power spectrum that affects the contrast transfer function (CTF) parameters determined from the power spectrum \cite{Wade1992brief,mindell2003accurate}. The CTF is described by the equation:

\begin{equation}
\label{CTF}
CTF(f)=-w_{\mathit{phase}}\sin{(\chi)}-w_{\mathit{amp}}\cos{(\chi)} 
\end{equation}

\noindent where \(w_{amp}\) is the fractional amount of amplitude contrast (\(\sim \)0.07 for a cryo-EM image \cite{toyoshima1988contrast}). The parameter \(w_{phase}\), the fractional amount of phase contrast, is given by  \(\sqrt{1-w_{amp}^{2}}\), and 


\begin{equation}
\label{chi}
\chi  =\frac{\pi \lambda}{f^2}\cdot \Delta z - \frac{\pi\lambda^3}{f^4}\cdot C_s
\end{equation}

\noindent where \(\lambda\) is the wavelength of electrons in \AA nstroms, \(f\) is the frequency in the Fourier transform in \AA nstroms\(^{-1}\), \(\Delta z\) is the defocus of the microscope in \AA nstroms, and \(C_s\) is the spherical aberration of the objective lens in \AA nstroms. For the resolutions currently of interest in biological cryo-EM (up to \(\sim\)2 \AA) and with the defocuses typically used (tens of thousands of \AA nstroms), the behaviour of the CTF is due almost entirely to the first term in equation \ref{chi}.  Consequently, a stretch of the image or contraction of the Fourier transform by a factor \(a\) will modify the apparent defocus in the image by a factor of \(1/a^{2}\). This phenomenon can be observed in Fig. 4A, which compares, for a 200 kV microscope with a \(C_s\) of 2 mm, a 1D CTF with \(\Delta z=\)10,000 \AA \ (black line), a 1D CTF with \(\Delta z=\)10,000 \AA \ but with the x-axis contracted by a factor of 1.1 \(\times\) (broken blue line), and a 1D CTF with \(\Delta z=\)10,000/\(1.1^2\) \AA = 8,264 \AA \ (red line).\\

\noindent For a 2D image two defocus values, \(\Delta z_{1}\) and \(\Delta z_{2}\), and an angle of astigmatism, \(\phi_{ast}\), are needed to describe the CTF (Fig. 4B). The angle of astigmatism \(\phi_{ast}\) is defined as the angle between the semi-axis of the Thon ring with defocus \(\Delta z_{1}\) and the \(k_{x}\)-axis of the Fourier transform. The parameters \(w_{phase}\), \(w_{amp}\), \(\lambda\), and \(C_s\) are microscope specific while \(\Delta z_1\), \(\Delta z_2\), and \(\phi_{ast}\) can change with each image. Equations \ref{CTF} and \ref{chi} may be used to describe a 2D CTF if one defines \(\Delta z\) in equation \ref{chi} as the effective defocus in the direction \(\phi\) as given by:

\begin{equation}
\Delta z=\frac{\Delta z_{1}+\Delta z_{2}+\cos (2 \cdot \left( {\phi-\phi_\mathit{ast}}\right))\left(\Delta z_{1}-\Delta z_{2}\right)}{2}
\end{equation}

\noindent During analysis of a dataset of cryo-EM images acquired on a FEI Company 200 kV FEG microscope equipped with a Gatan K2 Summit DDD, we plotted the measured contrast transfer function defocus parameters \(\Delta z_{1}\) versus \(\Delta z_{2}\) (Fig. 4C). Note that the choice of which defocus to define as  \(\Delta z_{1}\) and which to define as  \(\Delta z_{2}\) is arbitrary and can be reversed by changing \(\phi_\mathit{ast}\) by 180\textdegree. The plot showed an apparent systematic astigmatism: for randomly introduced astigmatism, one would expect points in the plot to fall on the line \(\Delta z_{1}\approx \Delta z_{2}\). In contrast, the points fall on a line that has a slope that indicated a constant astigmatism of approximately 2 \%, consistent with the measured magnification anisotropy of the the microscope. Plotting \(\Delta z_{1}\) versus \(\Delta z_{2}\) and looking for deviation from the line \(\Delta z_{1}=\Delta z_{2}\) with the form \(\Delta z_{1}\approx a\Delta z_{2}\) or \(\Delta z_{1}\approx \frac{1}{a}\Delta z_{2}\) presents a straightforward way of detecting anisotropic magnification in a microscope. However, if the anisotropic magnification factor \(a\) is small compared to the variance of the ratio of \(\Delta z_{1}/\Delta z_{2}\) due to errors in adjusting the microscope objective lens astigmatism, it will not be possible to detect anisotropic magnification in this way.\\

\begin{figure}[ht]
  \centering
  \includegraphics[width=0.9\textwidth]{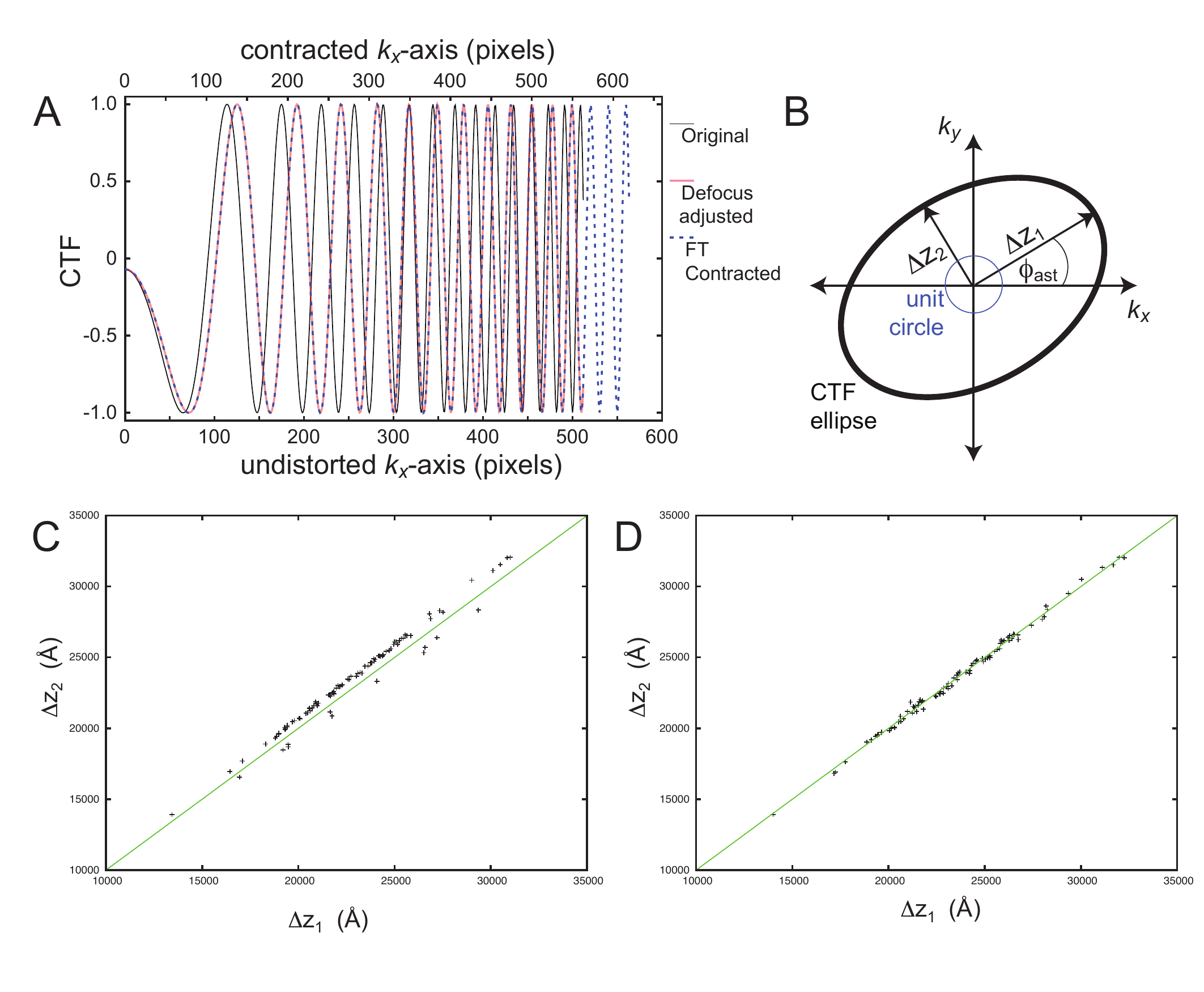}
  \caption{ \textbf{Effect of anisotropic magnification on CTF parameters A.} The effects of stretching or contracting a Fourier transform on the CTF can be corrected to high precision by adjusting the defocus parameter. \textbf{B,} Contrast transfer function (CTF) parameters that are unique for each image are the two defocus parameters \(\Delta z_1\), \(\Delta z_2\), and the angle of astigmatism \(\phi_{ast}\). \textbf{C,} A plot of \(\Delta z_1\) vs \(\Delta z_2\) for CTF parameters determined from images that suffer from magnification anisotropy show an apparent systematic astigmatism, with \(\Delta z_1\) consistently being larger (or smaller) than \(\Delta z_2\). Note that the ratio of \(\Delta z_1\) to \(\Delta z_2\)  can be reversed by redefining which defocus corresponds to which parameter and adding \(\frac{\pi}{2}\) to \(\phi_{ast}\). \textbf{C,} The apparent systematic astigmatism induced by anisotropic magnification can be removed by correcting CTF parameters as described in the text.}
\end{figure}

\noindent Correction of images for anisotropic magnification before measuring CTF parameters removes this systematic deviation from the \( \Delta z_{1}=\Delta z_{2}\) line. However, it is also possible to correct CTF parameters that have been measured from images with anisotropic magnification. It is advantageous to measure CTF parameters from images with magnification anisotropy and then correct the parameters for three reasons. First, computationally correcting whole images and movies, rather than individual particle images, could be computationally expensive. Second, storing computationally corrected whole images and movies places significant demand on data storage resources, which may already be strained by DDD data collection. Finally, correcting images for anisotropic magnification can introduce image artefacts that may decrease the accuracy with which CTF parameters can be calculated. Recovery of true parameters from images with a known amount of magnification anisotropy relies on representing the CTF as a distortion of the unit circle in Fourier space, \(k_{x}^2+k_{y}^2=1\), to an ellipse (Fig. 4B). The unit circle may be represented in matrix notation as 

\begin{equation}
\label{eqcirc}
\mathbf{k}^\mathrm{T}\mathbf{k}=1
\end{equation}

\noindent where  \(\mathbf{k}^\mathrm{T}=\begin{bmatrix}k_x & k_y\end{bmatrix}\) and \(\mathbf{k}=\begin{bmatrix}k_x \\ k_y\end{bmatrix}\). As seen in Fig. 4B, an ellipse can be used to represent the CTF parameters \(\Delta z_1\), \(\Delta z_2\), and \(\phi_\mathit{ast}\). Defining \(\mathbf{k}=\mathbf{E}_{CTF}\boldsymbol{\ell}\) where \(\mathbf{E}_{CTF}=\mathbf{R}_{CTF}\mathbf{S}_{CTF}\mathbf{R}^\mathrm{T}_{CTF}\) with 

\begin{equation*}
\mathbf{R}_{CTF}=
\begin{bmatrix}
\cos{\phi_\mathit{ast}} & -\sin{\phi_\mathit{ast}}\\
\sin{\phi_\mathit{ast}} & \cos{\phi_\mathit{ast}}\\
\end{bmatrix}
\text{and }
\mathbf{S}_{CTF}=
\begin{bmatrix}
\frac{1}{\sqrt{\Delta z_{1}}} & 0\\
0 & \frac{1}{\sqrt{\Delta z_{2}}}\\
\end{bmatrix}
\end{equation*}

\noindent and substituting into equation \ref{eqcirc} gives the equation of the ellipse as 

\begin{equation}
\label{eqelli1}
\boldsymbol{\ell}^\mathrm{T}\mathbf{E}_{CTF}^\mathrm{T}\mathbf{E}_{CTF}\boldsymbol{\ell}=1.
\end{equation}

\noindent \(\mathbf{R}_{CTF}^\mathrm{T}\) describes a rotation by the angle of astigmatism \(\phi_\mathit{ast}\) from the \(k_x\)-axis, \(\mathbf{S}_{CTF}\) describes a stretch or compression along the \(k_x\)- and \(k_y\)-axes by \(\Delta z_{1}\) and \(\Delta z_{2}\), and \(\mathbf{R}_{CTF}\) rotates the ellipse axes back to their correct angles. The magnitudes of the two defocus values (\(\Delta z_{1}\) and \(\Delta z_{2}\)) can be recovered by determining the two eigenvectors of \(\mathbf{E}_{CTF}^\mathrm{T}\mathbf{E}_{CTF}\). The angle of astigmatism \(\phi_\mathit{ast}\) can be calculated as the angle that the eigenvector representing \(\Delta z_{1}\) makes with the \(k_x\)-axis. By definition, the eigenvectors of a matrix \(\mathbf{Y}\) obey the equation \(\mathbf{Yx}=\lambda \mathbf{x}\). That is, upon multiplication by \(\mathbf{Y}\) the eigenvectors change only their magnitude, not their direction. Eigenvectors of a \(2 \times 2\) matrix can be conveniently obtained using algorithms in standard numerical analysis libraries such as \textit{LAPACK}, \textit{NumPy}, or in \textit{MATLAB}. As described earlier, magnification anisotropy further distorts the ellipse by the transformation \(\mathbf{E}_{ani}=\mathbf{R}_{ani}\mathbf{S}_{ani}\mathbf{R}^\mathrm{T}_{ani}\) where \(\mathbf{R}_{ani}\) describes a rotation by the angle of magnification anisotropy \(\theta_\mathit{ani}\) from the \(k_x\)-axis, and \( \mathbf{S}_{ani} \) describes a stretch or compression along the \(k_x\)-axis by the magnitude of magnification anisotropy \(a\), where \(\mathbf{R}_{ani}\) and \(\mathbf{S}_{ani}\) are the shown in \ref{anisotropy}. Defining \(\boldsymbol{\ell}=\mathbf{E}_{ani}\mathbf{m}\) and substituting this definition into equation \ref{eqelli1} gives the equation of the new ellipse as 

\begin{equation}
\label{eqelli2}
\mathbf{m}^\mathrm{T}\mathbf{E}_{ani}^\mathrm{T}\mathbf{E}_{CTF}^\mathrm{T}\mathbf{E}_{CTF}\mathbf{E}_{ani}\mathbf{m}=1.
\end{equation}

\noindent The eigenvectors of \(\mathbf{E}_{ani}^\mathrm{T}\mathbf{E}_{CTF}^\mathrm{T}\mathbf{E}_{CTF}\mathbf{E}_{ani}\) correspond to the apparent CTF parameter values, \(\Delta z^\prime_{1}\), \(\Delta z^\prime_{2}\), and \(\phi^\prime_\mathit{ast}\), measured from the power spectra of images where there is aniostropic magnification. From these values and values of \(a\) and \(\theta_\mathit{ani}\) measured from a powder diffraction pattern (see above) it is possible to calculate the true values of \(\Delta z_{1}\), \(\Delta z_{2}\), and \(\phi_\mathit{ast}\) from the eigenvectors of 

\begin{equation}
\mathbf{E}_{CTF}^\mathrm{T}\mathbf{E}_{CTF}=\mathbf{E}_{ani}^{-\mathrm{T}}\mathbf{E}_{CTF}^{\prime\mathrm{T}}\mathbf{E}_{CTF}^{\prime}\mathbf{E}_{ani}^{-1}
\end{equation}

\noindent where \(\mathbf{E}_{CTF}^\prime=\mathbf{R_\mathit{CTF^\prime}^\mathrm{T}S_\mathit{CTF^\prime}R_\mathit{CTF^\prime}}\) and \(\mathbf{E}_{ani}^{-1}=\mathbf{R}_{ani}^\mathrm{T}\mathbf{S}_{ani}^{-1}\mathbf{R}_{ani}\). The matrices \(\mathbf{R}_{CTF^\prime}\), \(\mathbf{R}_{CTF^\prime}^\mathrm{T}\), and \(\mathbf{S}_{CTF^\prime}\) are the same as \(\mathbf{R}_{CTF}\), \(\mathbf{R}_{CTF}^\mathrm{T}\), and \(\mathbf{S}_{CTF}\), but use the apparent CTF parameters measured from distorted images rather than the true CTF parameters, and 

\begin{equation*}
\mathbf{S}_{ani}^{-1}=
\begin{bmatrix}
1/a & 0\\
0 & 1\\
\end{bmatrix}.
\end{equation*}

\noindent This approach for correcting CTF parameters measured from images with magnification anisotropy was implemented in a standalone program that operates on \textit{Relion} .star files \cite{scheres2012relion}. As can be seen in Fig. 4D, the method brings points in a plot of \(\Delta z_1\) versus \(\Delta z_2\) back to the \(\Delta z_1=\Delta z_2\) line. The method is equivalent to correcting an image for magnification anisotropy and subsequently measuring CTF parameters from its power spectrum.

\section{Discussion}

\noindent As seen in Fig. 4C, a plot of \(\Delta z_1\) versus \(\Delta z_2\) from a dataset of micrographs can reveal the presence of anisotropic magnification in the microscope. If all of the images in a dataset were obtained from a single EM session where the objective lens stigmator was adjusted once, it would not be unusual to find the points of this plot fall off the line \(\Delta z_1=\Delta z_2\). However, if the deviation of points from the line is due to anisotropic magnification, the positions of the points should have the form \(\Delta z_1\approx a\Delta z_2\) or \(\Delta z_1\approx \frac{1}{a}\Delta z_2\). Deviation due a constant astigmatism should result in point positions with the form \(\Delta z_1\approx \Delta z_2+x\), where \(x\) is the the amount of astigmatism. Once anisotropic magnification has been detected, the amount and extent of anisotropy can be measured precisely using a diffraction standard such as the thallous chloride shown above. The interpolation scheme proposed here is not entirely free of artefacts. Consequently, it would be better to correct anisotropic magnification with improved electron optics, rather than correct the effects of anisotropic magnification computationally. If correction by interpolation must be performed, the best approach would be to incorporate the correction into the 3D map calculation software in order to avoid performing interpolation twice with the same images.

\section{Software availability}
All of the original software described above is available from https://sites.google.com/site/rubinsteingroup/direct-detector-distortion

\section{Acknowledgements}
We are grateful to Alexis Rohou and Timothy Grant from Nikolaus Grigorieff's laboratory, who quickly put us on the correct track when we noticed the apparent systematic astigmatism in plots of \(\Delta z_1\) versus \(\Delta z_2\). We thank Richard Henderson for critical comments on the manuscript. JZ was supported by a Doctoral Postgraduate Scholarship from the National Science and Engineering Council of Canada and a Mary Gertrude l'Anson Scholarship. JLR was supported by a Canada Research Chair. This work was funded by NSERC grant 401724-12 to JLR. 

\section{Author contributions}
JZ noticed the systematic error in CTF parameters. SB Acquired the images of thallous chloride. JLR analyzed the images of thallous chloride to measure the anisotropy of magnification, and wrote the programs to correct images for anisotropic magnification in real space and in Fourier space. JLR, JZ, and MAB conceived of the approach to correct CTF parameters measured from images with anisotropic magnification and JZ wrote the program. JLR made the figures and JLR and JZ wrote the manuscript.

\bibliographystyle{elsarticle-num}
\bibliography{./Bibliography}

\end{document}